\documentclass[namedreferences]{solarphysics}

\usepackage[hyperref,optionalrh]{spr-sola-addons} 
\usepackage{graphicx}        
\usepackage{color}           
\usepackage{breakurl}        

\chardef\us=`\_

\begin{document}

\begin{article}
\begin{opening}

\title{Dependence of great geomagnetic storm intensity ($\Delta$SYM-H$\le$-200 nT) on associated solar wind parameters}

\author[addressref={aff1}]{\fnm{Ming-Xian}~\lnm{Zhao}}
\author[addressref={aff1,aff2},corref,email={Legm@cma.gov.cn}]{\inits{G.-M.}\fnm{Gui-Ming}~\lnm{Le}}
\author[addressref={aff3}]{\fnm{Qi}~\lnm{Li}}
\author[addressref={aff2}]{\fnm{Gui-Ang}~\lnm{Liu}}
\author[addressref={aff1}]{\fnm{Tian}~\lnm{Mao}}

\address[id=aff1]{Key Laboratory of Space Weather, National Center for Space Weather, China Meteorological Administration, Beijing, 100081, P.R. China}
\address[id=aff2]{School of Physics Science and Technology, Lingnan Normal University, Zhanjiang, 524048, P.R. China}
\address[id=aff3]{Institute of Geophysics, China Earthquake Administration, Beijing, 100081, P. R. China}

\runningauthor{M.-X. Zhao et al.}
\runningtitle{Dependence of great Geomagnetic storms on Solar Wind Parameters}

\begin{abstract}
We use $\Delta$SYM-H to capture the variation in the SYM-H index during the main phase of a geomagnetic storm. We define great geomagnetic storms as those with $\Delta$SYM-H $\le$ -200 nT. After analyzing the data that were not obscured by solar winds, we determined that 11 such storms occurred during solar cycle 23. We calculated time integrals for the southward interplanetary magnetic field component I(B$_s$), the solar wind electric field I(E$_y$), and a combination of E$_y$ and the solar wind dynamic pressure I(Q) during the main phase of a great geomagnetic storm. The strength of the correlation coefficient (CC) between $\Delta$SYM-H and each of the three integrals I(B$_s$) (CC = 0.74), I(E$_y$) (CC = 0.85), and I(Q) (CC = 0.94) suggests that  Q, which encompasses both the solar wind electric field and the solar wind dynamic pressure, is the main driving factor that determines the intensity of a great geomagnetic storm. The results also suggest that the impact of B$_s$ on the great geomagnetic storm intensity is much more significant than that of the solar wind speed and the dynamic pressure during the main phase of associated great geomagnetic storm. How to estimate the intensity of an extreme geomagnetic storm based on solar wind parameters is also discussed.
\end{abstract}
\end{opening}

\section{Introduction} \label{sec:intro}

A geomagnetic storm is the result of the sustained interaction between solar winds with the southward magnetic field and Earth's magnetic field. Previous works have explored the effect of different solar wind parameters on the intensity of an associated geomagnetic storm by calculating the correlation coefficients (CCs) between the peak value of various solar wind parameters and the minimum Dst index for the associated geomagnetic storm \citep[e.g.][]{Echer2008GRL, Echer2008JGR, Choi2009, Kane2005, Kane2010, Ji2010, Richardson2011, WuCC2002, WuCC2016, Meng2019, Lawrance2020}. However, these CC values have no physical meaning \citep{Le2020}.

\cite{WangYM2003} proposed that the geomagnetic storm intensity is largely unaffected by the solar wind density or the dynamic pressure, and that it is only a function of the interplanetary dawn-dusk electric field (termed as the solar wind electric field in this study). \cite{Echer2008GRL} determined that the CC between the time integral of the solar wind electric field during the main phase of the super geomagnetic storm intensity and the minimum of Dst index is equal to 0.62. \cite{Balan2014, Balan2017} explored the relationship between super geomagnetic storms, the sudden high enhancement in the solar wind speed, and the southward magnetic field at the leading edge of the associated coronal mass ejection (CME). Based on the work of \cite{Burton1975}, \cite{Kumar2015} estimated the magnitude of the interplanetary electric fields responsible for historical geomagnetic storms. \cite{Liu2014APJ, Liu2014NC, Liu2020} evaluated an extreme geomagnetic storm intensity only based on solar wind electric field, without considering the effect of the solar wind dynamic pressure on the extreme geomagnetic storm intensity. \cite{Xue2005} identified the interplanetary sources that were responsible for the great geomagnetic storms (Dst $\le$ -200 nT) that occurred during the solar maximum (2000 - 2001) and quantified the linear fit between Dst and both the solar wind electric field and the storm duration with the solar wind dynamic pressure making little contribution to the storm intensity. These published works provide valuable insight into geomagnetic storms, but largely ignore the possible contributions made by the solar wind density or the solar wind dynamic pressure.

Case studies \citep{Kataoka2005, Cheng2020}, global MHD simulations \citep{Lopez2004}, and an impulse response function model \citep{Weigel2010} suggest that the solar wind density is an important parameter that modulating the transfer of solar wind energy to the magnetosphere during the main phase of a storm.

The development of a geomagnetic storm depends on the ring current injection term, Q, and the decay term. Q is either implemented as a linear function of the solar wind electric field \citep{Burton1975, Fenrich1998, OBrien2000JGR}, or as a function of both the solar wind electric field and the solar wind dynamic pressure \cite{WangCB2003}. A recent study has shown that it is more appropriate to apply the definition of Q that includes the solar wind dynamic pressure for major geomagnetic storms (Dst $\le$ -100 nT) \citep{Le2020}.

\cite{Le2020} found that the time integrals for the southward interplanetary magnetic field component I(B$_s$), the solar wind electric field I(E$_y$), and a combination of E$_y$ and the solar wind dynamic pressure I(Q) during the main phase of the major geomagnetic storm make small, moderate, and crucial contributions to the intensity of the major geomagnetic storm, respectively. Great geomagnetic storms ($\Delta$SYM-H $\le$ -200 nT) are much stronger than major geomagnetic storms (Dst$_{min}$ $\le$ -100 nT). To determine whether a similar statistical trend exists for great geomagnetic storms, we calculated the CCs between $\Delta$SYM-H and these three time integrals for great geomagnetic storms.

The stronger the geomagnetic storm, the worse the space weather and the greater the harm. How to estimate the intensity of an extreme geomagnetic storm? Researchers \citep[e.g.][]{Liu2014APJ, Liu2014NC, Liu2020} usually estimated the intensity of an extreme geomagnetic storm by using  the empirical formulas found by \cite{Burton1975, Fenrich1998, OBrien2000JGR}. However, the empirical formula found by Wang, Zhao and Lin (2003) was seldom used to estimate the intensity of an extreme geomagnetic storm. Which empirical formula is better to describe the relation between solar wind parameters and the intensity of an extreme storm? To answer the question, we will discuss which empirical formula is the better one. The data analysis, discussion, and summary for this study are presented in Section \ref{sec:data}, Section \ref{sec:discussion}, and Section \ref{sec:summary}, respectively.

\section{Data analysis} \label{sec:data}

\subsection{Solar Wind Data and Geomagnetic Storm Data} \label{subsec:sw-gm-data}

The SYM-H index was obtained from the World Data Center for Geomagnetism in Kyoto (\url{http://wdc.kugi.kyoto-u.ac.jp/aeasy/index.html}). In this study, our data set consists of solar wind data recorded by the Advanced Composition Explorer (ACE) (\url{ftp://mussel.srl.caltech.edu/pub/ace/level2/magswe}) from 1998 to 2006 with a time resolution of 64 seconds.

\subsection{The criteria for a great geomagnetic storm}

Seventeen geomagnetic storms with a minimum of Dst $\le$ -200 nT occurred during solar cycle 23. However, the main phases of five of those great geomagnetic storms coincided with a data gap caused by the solar wind. The SYM-H index can be treated as a high time resolution Dst index \citep{Wanliss2006}, we used $\Delta$SYM-H to represent the variation in SYM-H during the main phase of a geomagnetic storm. In this study, we define storms with $\Delta$SYM-H $\le$ -200 nT as great geomagnetic storms. The geomagnetic storm that occurred on November 9$^{th}$ and 10$^{th}$ in 2004, with $\Delta$SYM-H = -165 nT during the main phase of the geomagnetic storm, does not meet the criteria for a great geomagnetic storm; as such, we do not include it in our data set. Because the variation in SYM-H during the main phase of the geomagnetic storm on September 25$^{th}$ in 1998 is equal to -177 nT, this storm is also not included in our data set. The minimum Dst value for a major geomagnetic storm that occurred on October 21$^{st}$ 2001 is -184 nT. However, because $\Delta$SYM-H = -217 nT during the main phase of the storm, we treat this storm as a great geomagnetic storm. Our final data set consists of the solar wind parameters for 11 great geomagnetic storms that occurred during solar cycle 23.

\subsection{The calculation of the solar wind parameters}

I(B$_s$), I(E$_y$), and I(Q) represent the time integrals of the southward component of interplanetary magnetic field (IMF), the solar wind electric field, and the ring current injection term \citep{WangCB2003} during the main phase of the associated great geomagnetic storm, respectively. These integrals are defined as
\begin{equation}\label{eq-01}
I(B_s) = \int_{t_{start}}^{t_{end}}B_z \mathrm{d}t
\end{equation}
\begin{equation}\label{eq-02}
I(E_y) = \int_{t_{start}}^{t_{end}}V_{sw}B_z \mathrm{d}t
\end{equation}
\begin{equation}\label{eq-03}
I(Q) = \int_{t_{start}}^{t_{end}}Q \mathrm{d}t
\end{equation}
where $t_{start}$ and $t_{end}$ are the start and the end times of the main phase of a great geomagnetic storm, respectively. The Q variable in Equation (\ref{eq-03}) was found by \cite{WangCB2003}, which is defined as
\begin{equation}\label{eq-04}
  Q = \left\{ \begin{array}{ll}
                0 & V_{sw}B_s \le 0.49 mV/m \\
                -4.4(V_{sw}B_s-0.49)\left(P_k/3\right)^{1/3} &  V_{sw}B_s > 0.49 mV/m
              \end{array} \right.
\end{equation}
where $P_k$ is the solar wind dynamic pressure.

The geomagnetic storm main phase is defined as a period from the moment when SYM-H starts to decrease to the moment when SYM-H reaches its lowest value. The start time of the main phase of a geomagnetic storm corresponds to the moment when z-component of interplanetary magnetic field starts to become negative. $\Delta$SYM-H represents the difference in the storm magnitude between the beginning and the end of the main phase of a great geomagnetic storm.


According to Equation listed below,
\begin{equation}\label{eq-05}
  \mathrm{d}\textrm{Dst}^*/\mathrm{\textrm{d}}t = \textit{Q(t)}-\textrm{Dst}^*/{\tau},
\end{equation}

Equation (5) can be written as below,

\begin{equation}\label{eq-07}
  \int_{\textit{t}_\textit{s}}^{\textit{t}_\textit{e}}\mathrm{\textit{d}}(\textrm{SYM-H}^*)=\int_{\textit{t}_\textit{s}}^{\textit{t}_\textit{e}} (\textit{Q(t)}-\textrm{SYM-H}^*/\tau)\mathrm{\textit{d}}\textit{t}.
\end{equation}
Where t$_{s}$ and t$_{e}$ are the start and the end time of the main phase of the associated storm. Assuming that the time integral of injection term is much larger than the time integral of the decay term during the main phase of a great geomagnetic storm, then we have
\begin{equation}\label{eq-08}
  \int_{\textit{t}_s}^{\textit{t}_e}\mathrm{\textit{d}}(\textrm{SYM-H}^*)\simeq \int_{\textit{t}_s}^{\textit{t}_e} \textit{Q(t)}\mathrm{\textit{d}}t
\end{equation}

$\left( {7.6\sqrt {{{\left. {{\textit{P}_{k}}} \right|}_{{t_s}}}}  - 7.6\sqrt {{{\left. {{\textit{P}_{k}}} \right|}_{{t_e}}}} } \right)$ is much smaller than $\triangle$SYM-H for a great geomagnetic storm.
Finally, we get the formula listed bellow,
\begin{equation}\label{eq-07}
  \Delta {\textrm{SYM-H}} \simeq \int_{t_{s}}^{t_{e}}Q(t) \mathrm{d} t = I(Q)
\end{equation}

After determining the start and the end times of the main phase of the great geomagnetic storm, we calculate $\Delta$SYM-H, identify the interplanetary source responsible for that $\Delta$SYM-H, and calculate the corresponding solar wind parameters I(B$_s$), I(E$_y$), and I(Q).

After determining the start and the end times of the main phase of the great geomagnetic storm, we calculate $\Delta$SYM-H, identify the interplanetary source responsible for that $\Delta$SYM-H, and calculate the corresponding solar wind parameters I(B$_s$), I(E$_y$), and I(Q).

\begin{figure}
    \centering
    \includegraphics[angle=90,width=0.9\textwidth]{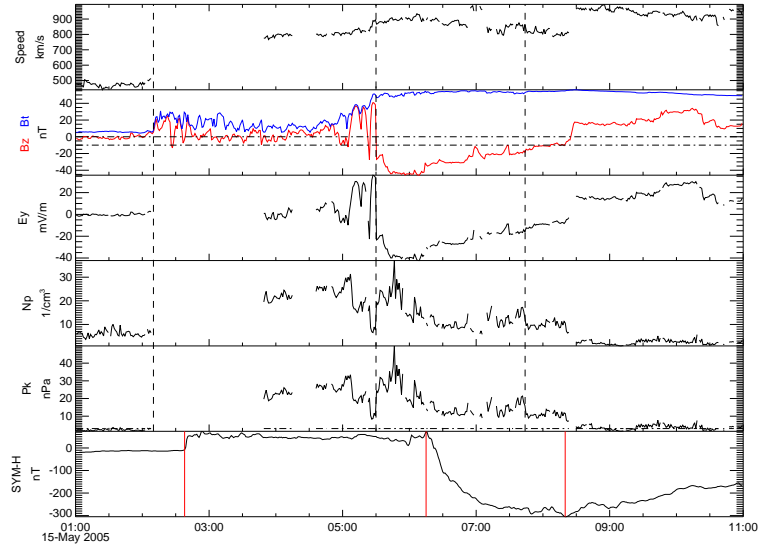}
    \caption{ACE spacecraft observations on May 15$^{th}$, 2005. From top to bottom, the panels represent the solar wind speed, the total IMF (B$_t$) (blue line) and the z-component of the IMF (B$_z$) (red line), the solar wind electric field (E$_y$), the proton density (N$_p$), the solar wind dynamic pressure (P$_k$), and the SYM-H values. The two horizontal dashed lines in the second panel denote 0 nT and -10 nT, respectively. The horizontal dashed line in the fifth panel denotes 3 nPa.}
    \label{fig-01}
\end{figure}

For example, let us examine the great geomagnetic storm that occurred on May 15$^{th}$ in 2005 (Fig. \ref{fig-01}). The ACE spacecraft recorded an interplanetary shock at 02:05 UT on May 15$^{th}$ in 2005 (the first vertical dashed line in Fig. \ref{fig-01}). The shock reached the magnetosphere at 02:38 UT and caused a sudden storm (the first vertical solid red line in Fig. \ref{fig-01}). The main phase of the storm, which is the period between the second and third vertical solid red lines in Figure \ref{fig-01}, has a $\Delta$SYM-H value of -350 nT. Solar wind between the second and third vertical dashed lines in Figure \ref{fig-01} is the interplanetary source responsible for the main phase of the storm. The I(B$_s$), I(E$_y$), and I(Q) values for the main phase of the storm are -3723.95 nT$\cdot$min, -3292.92 mV$\cdot$m$^{-1}\cdot$min, and -35045.2 mV$\cdot$m$^{-1}\cdot$nPa$\cdot$min, respectively.

\begin{figure}
  \centering
  \includegraphics[angle=90,width=0.9\textwidth]{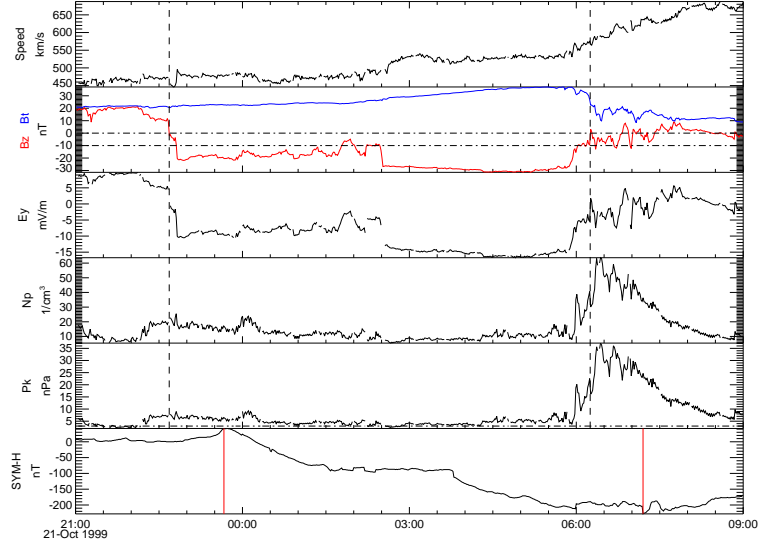}
  \caption{ACE spacecraft observations on October 21$^{st}$ and 22$^{nd}$ in 1999. The panels and lines are identical to those shown in Figure 1.}
  \label{fig-02}
\end{figure}

The calculations for a second storm, which occurred on October 21$^{st}$ to 22$^{nd}$ in 1999, are shown in Figure \ref{fig-02}. The main phase of the storm is the period between the first and second vertical red lines, and it was caused by the solar wind between the first and second vertical dashed lines in Figure \ref{fig-02}. During the main phase of the storm, $\Delta$SYM-H = -269 nT, and the I(B$_s$), I(E$_y$), and I(Q) values are -9183.60 nT$\cdot$min, -4647.12 mV$\cdot$m$^{-1}\cdot$min, and -24798.4 mV$\cdot$m$^{-1}\cdot$nPa$\cdot$min, respectively.

\subsection{The results}

After determining the I(B$_s$), (I(E$_y$), I(Q), and $\Delta$SYM-H values for each great geomagnetic storm, we calculated the CC values between $\Delta$SYM-H and each of the three time integrals: CC(I(B$_s$), $\Delta$SYM-H) = 0.74, CC(I(E$_y$), $\Delta$SYM-H) = 0.85, and CC(I(Q), $\Delta$SYM-H) = 0.94.


  \begin{figure}    
   \centerline{\hspace*{0.015\textwidth}
               \includegraphics[width=0.515\textwidth,clip=]{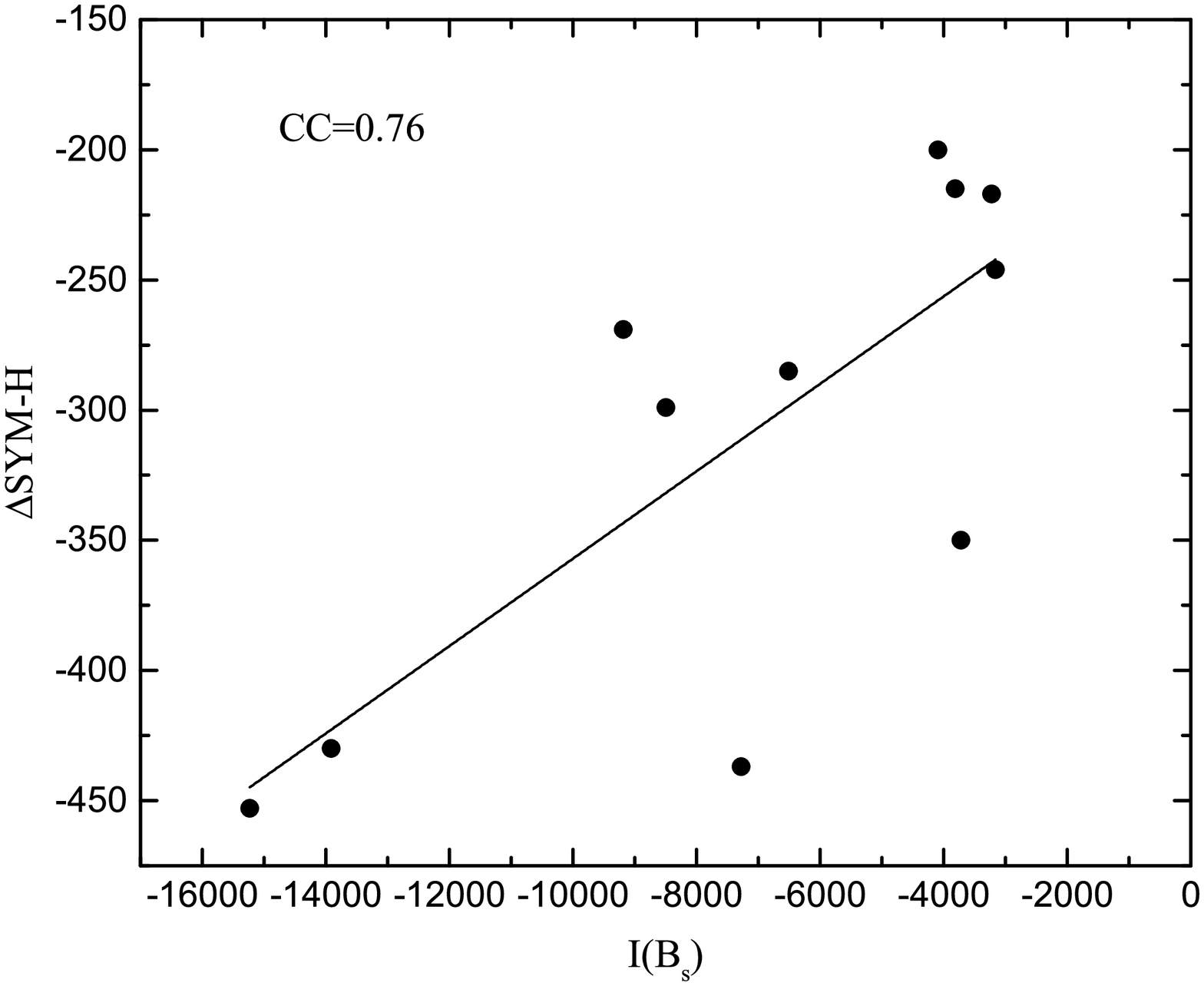}
               \hspace*{-0.03\textwidth}
               \includegraphics[width=0.515\textwidth,clip=]{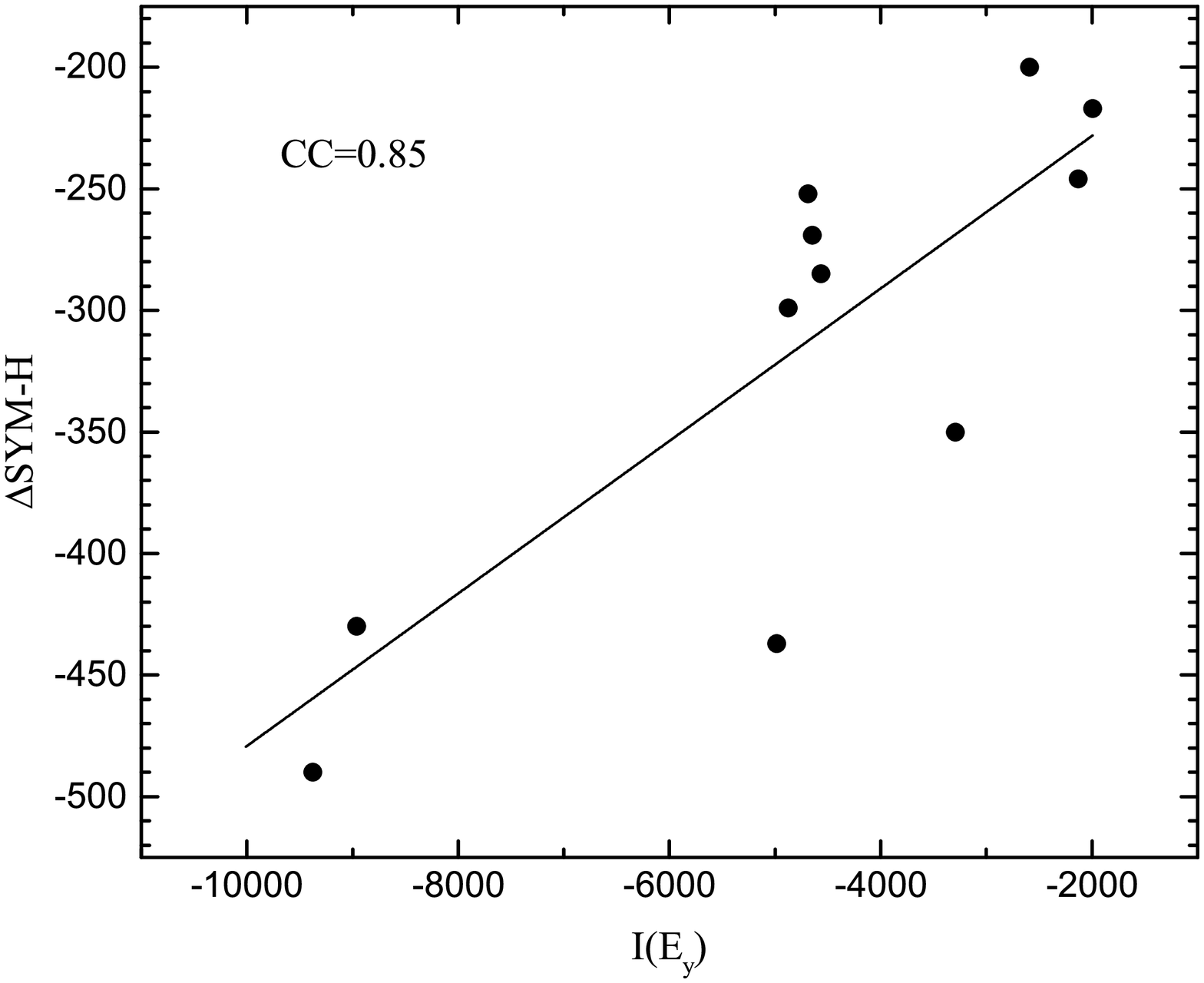}
              }
     \centerline{     
      \hspace{0.24 \textwidth}  \color{black}{(a)}
      \hspace{0.46 \textwidth}  \color{black}{(b)}
         \hfill}
%
   \centerline{\hspace*{0.015\textwidth}
               \includegraphics[width=0.515\textwidth,clip=]{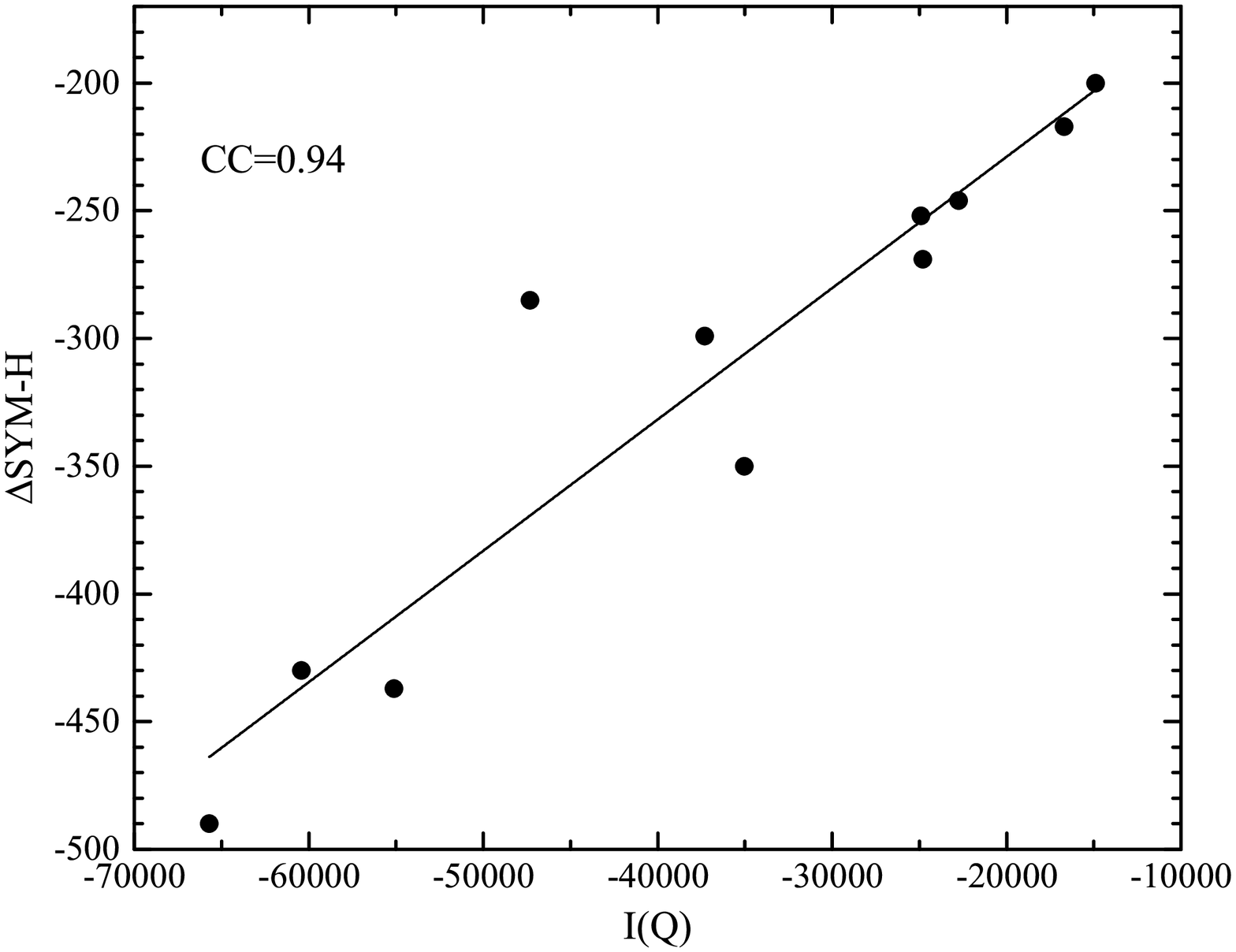}
              }
     \centerline{     
      \hspace{0.48 \textwidth} \color{black}{(c)}
         \hfill}

\caption{Statistical analyses of the relationships between $\Delta$SYM-H and (a) I(B$_s$), (b) I(E$_y$), and (c) I(Q).
        }
   \label{fig-03}
   \end{figure}

\section{Discussion} \label{sec:discussion}

A previous estimation of CC(I(E$_y$), Dst$_{\textrm{min}}$) = 0.62 \citep{Echer2008GRL} for super geomagnetic storms (Dst$_{\textrm{min}}$$\le$ -250 nT) is much lower than our CC(I(E$_y$), $\Delta$SYM-H) value of 0.85 for great geomagnetic storms. The difference in these values arises due to the difficulty in determining the start and end time of a geomagnetic storm precisely using the Dst index. The time resolution of the Dst index is one hour; the SYM-H index has a much finer time resolution, which allows for more exact determination of the start and end time of the main phase of a geomagnetic storm, and then the interplanetary source responsible for the geomagnetic storm main phase can be determined exactly. {SYM-H$_{\textrm{min}}$ is not equal to $\Delta$SYM-H for some storms. For example, SYM-H$_{\textrm{min}}$ for the geomagnetic storm on 9-10, November 2004 is -214 nT, while $\Delta$SYM-H is -165 nT, which is much higher than SYM-H$_{\textrm{min}}$. Equation (\ref{eq-07}) tell us that CC(I(Q), $\Delta$SYM-H) is more reasonable than CC(I(Q), SYM-H$_{\textrm{min}}$). CC(I(E$_y$), $\Delta$SYM-H) is also more reasonable than CC(I(E$_y$), SYM$_{\textrm{min}}$). These should be also the reason why CC(I(E$_y$), $\Delta$SYM-H) in the present study is larger than CC(I(E$_y$), Dst$_{\textrm{min}}$) in the article by \cite{Echer2008GRL}.}

CC(I(B$_s$), $\Delta$SYM-H) for great geomagnetic storms is 0.74, while CC(I(B$_s$), SYM-H$_{\textrm{min}}$) for major geomagnetic storms is o.33 \citep{Le2020}, indicating that there is a big difference between the CC(I(B$_s$), $\Delta$SYM-H) for great geomagnetic storms and {CC(I(B$_s$), Dst$_{\textrm{min}}$)} for major geomagnetic storms. We attribute this differences to the fact that the B$_s$ value for a great geomagnetic storm is much larger than that of a major geomagnetic storm, {and CC(I(B$_s$), $\Delta$SYM-H) is more reasonable than CC(I(B$_s$), SYM-H$_{\textrm{min}}$)}. Because $\textrm{E}_\textrm{y} = \textrm{V}_{\textrm{sw}}\textrm{B}_\textrm{s} $, CC(I(B$_s$), $\Delta$SYM-H) and CC(I(E$_y$), $\Delta$SYM-H) are 0.74 and 0.85 respectively, indicating that the contribution to the great geomagnetic intensity made by the solar wind speed is much smaller than southward component of IMF. The comparison between CC(I(B$_s$), $\Delta$SYM-H) and CC(I(Q), $\Delta$SYM-H) implies that the contributions to the great geomagnetic intensity made by the solar wind speed and dynamic pressure are much lower than southward component of IMF. Based on our CC values, we infer that B$_s$ contribution to the great geomagnetic storm intensity is much more significant than those of the solar wind speed and the dynamic pressure.

CC(I(Q), $\Delta$SYM-H) is larger than CC(I(E$_y$), $\Delta$SYM-H), suggesting that the ring current injection term Q that includes both the solar wind electric field and the solar wind dynamic pressure is more accurate than the Q definition that only depends on the solar wind electric field for great geomagnetic storms.

For Q defined in Equation (\ref{eq-04}), $\textrm{E}_\textrm{y}$ should be much larger than 0.49 mV/m  during the main phase of a great geomagnetic storm. As such, Q can be written as
\begin{equation}\label{eq-08}
 Q = -4.4 (E_y-0.49)(P_k/3)^{1/3}
\end{equation}

The averaged P${_k}$ for the great geomagnetic storms ranged from 5.3 nPa to 26.5 nPa, indicating that 1.2$<$$(P_k/3)^{1/3}$$<$2.1. The value of averaged E$_y$ for great geomagnetic storms ranged from 7 mV/m to 26 mV/m, much larger than that of averaged $(P_k/3)^{1/3}$. {According to Equation (\ref{eq-08}), the impact of E$_y$ on Q is much larger than $(P_k/3)^{1/3}$}. This should be the reason why CC(I(Q), $\Delta$SYM-H) is slightly larger than CC(I(E$_y$), $\Delta$SYM-H), namely that the difference between I(Q) and I(E$_y$) is small for the great geomagnetic storms. If I(Q) is very close to I(E$_y$) for extreme geomagnetic storm, then the intensity of an extreme geomagnetic storm can be estimated by I(E$_y$). However, what is the real situation? Here, we give a real case. The solar wind dynamic pressure during the some part of the main phase of the extreme geomagnetic storm shown in Figure 4 of the article by \cite{Liu2020} exceeded 700 nPa. The solar wind dynamic pressure is very big and should have a significant effect on the associated extreme storm intensity according to Equation (\ref{eq-08}), implying that the extreme magnetic storm intensity shown in Figure 4 of the article by \cite{Liu2020} was greatly underestimated. Anyway, CC(I(Q), $\Delta$SYM-H) is always larger than CC(I(E$_y$), $\Delta$SYM-H) for great geomagnetic storms and even stronger geomagnetic storms, implying that it is more accurate to estimate the intensity of an extreme geomagnetic storm by I(Q) than  by I(E$_y$).

\section{Summary} \label{sec:summary}

Our CC values that capture the statistical relationship between $\Delta$SYM-H and I(B$_s$), I(E$_y$), and I(Q), where Q is a combination of E$_y$ and the solar wind dynamic pressure, for great geomagnetic storms ($\Delta$SYM-H $\le$ -200 nT) are equal to 0.74, 0.85, and 0.94, respectively. With the strength of the correlation between $\Delta$SYM-H and I(Q), we infer that Q is the most important solar wind parameter in the determination of the intensity of a great geomagnetic storm. Furthermore, our results imply that it is more accurate to use the ring current injection term definition proposed by \cite{WangCB2003} than it is to define Q as a linear function of the solar wind electric field when it comes to assessing the intensity of a great geomagnetic storm and even stronger geomagnetic storm. The statistical results also suggest that B$_s$ makes much more contribution to the intensities of great geomagnetic storms that happened in Solar Cycle 23 than solar wind speed and dynamic pressure.

\begin{acks}
We thank LetPub (\url{www.letpub.com}) for its linguistic assistance during the preparation of this manuscript. We thank the ACE SWEPAM instrument team and the ACE Science Center for providing the ACE data. We thank Center for Geomagnetism and Space Magnetism, Kyoto University, for providing SYM-H index. We also thank Institute of Geophysics, China Earthquake Administration for providing sudden storm commence data. This work is supported by the National Natural Science Foundation of China (Grant No. 41774085, 41074132, 41274193, 41474166, 41774195, 41874187)
\end{acks}


\bibliographystyle{spr-mp-sola}
\bibliography{solphy2020.bib}

\end{article}

\end{document}